\documentclass{ecai} 

\usepackage{latexsym}
\usepackage{amssymb}
\usepackage{amsmath}
\usepackage{amsthm}
\usepackage{booktabs}
\usepackage{enumitem}
\usepackage{graphicx}
\usepackage{color}
\usepackage{multicol}
\usepackage{multirow}

\newcommand{\BibTeX}{B\kern-.05em{\sc i\kern-.025em b}\kern-.08em\TeX}

\begin{document}


\begin{frontmatter}


\paperid{6378} 


\title{Token Pruning in Audio Transformers: Optimizing Performance and Decoding Patch Importance}


\author[A]{\fnms{Taehan}~\snm{Lee}\orcid{0009-0009-3576-2770}}
\author[A]{\fnms{Hyukjun}~\snm{Lee}\orcid{0000-0003-2981-0800}\thanks{Corresponding Author. Email: hyukjunl@sogang.ac.kr}}

\address[A]{Department of Computer Science and Engineering, Sogang University, South Korea}


\begin{abstract}
Vision Transformers (ViTs) have achieved state-of-the-art performance across various computer vision tasks, but their high computational cost remains a challenge. Token pruning has been proposed to reduce this cost by selectively removing less important tokens. While effective in vision tasks by discarding non-object regions, applying this technique to audio tasks presents unique challenges, as distinguishing relevant from irrelevant regions in time-frequency representations is less straightforward. In this study, for the first time, we applied token pruning to ViT-based audio classification models using Mel-spectrograms and analyzed the trade-offs between model performance and computational cost: TopK token pruning can reduce MAC operations of AudioMAE and AST by 30-40\%, with less than a 1\% drop in accuracy.
Our analysis reveals that while high-intensity or high-variation tokens contribute significantly to model accuracy, low-intensity or low-variation tokens also remain important when token pruning is applied; pruning solely based on the intensity or variation of signals in a patch leads to a noticeable drop in accuracy. We support our claim by measuring high correlation between attention scores and these statistical features and by showing retained tokens consistently receive distinct attention compared to pruned ones.
We also show that AudioMAE retains more low-intensity tokens than AST. This can be explained by AudioMAE’s self-supervised reconstruction objective, which encourages attention to all patches, whereas AST’s supervised training focuses on label-relevant tokens.

\end{abstract}

\end{frontmatter}


\section{Introduction}

The Vision Transformer (ViT)~\cite{vit} has achieved various SOTA (state-of-the-art) performance in many downstream tasks. As the Transformer \cite{transformer} has data-agnostic characteristics and audio can be represented as 2D data using Mel-spectrograms, previous studies \cite{dftat, htsat, eat, ast, ssast, audiomae, atst, m2d} have shown the applicability of ViT to audio downstream tasks.
To reduce the high computational demands of Transformer-based models, token reduction methods have been proposed, as the number of tokens is a quadratic factor in both time and memory complexity. In image classification tasks, various token pruning methods based on attention scores~\cite{ats, evit, romance} or DNN predictors~\cite{spvit, dynamicvit} have demonstrated favorable trade-offs between accuracy and computational cost. 
\begin{figure}[ht]
  \centering
  \includegraphics[width=\linewidth]{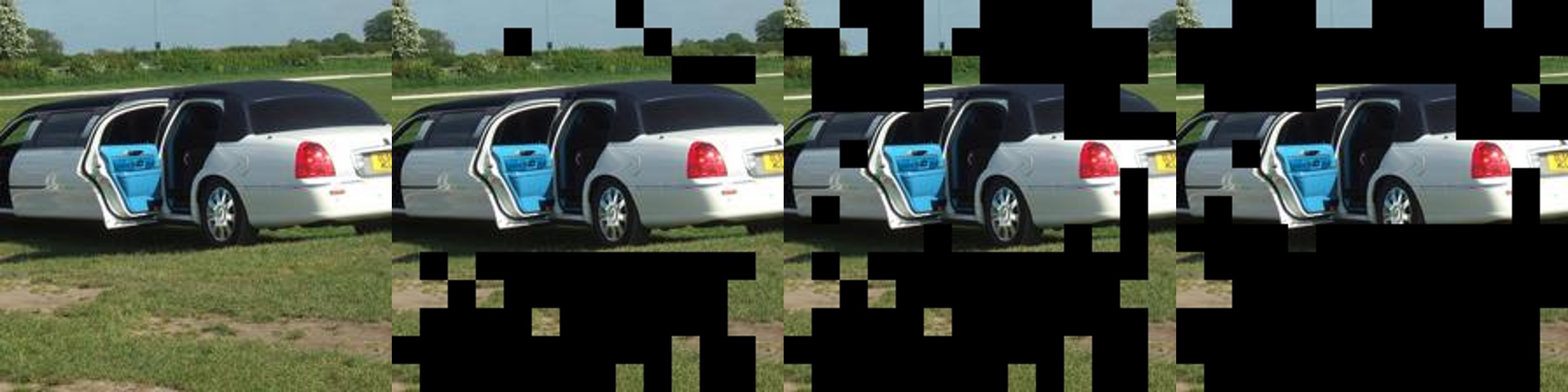}
  \caption{Token pruning patterns in the image classification ViT~\cite{evit}. The model gradually prunes tokens less relevant to object being classified.}
  \label{fig:evit_visualization}
\end{figure}
As shown in Fig.~\ref{fig:evit_visualization}, these methods gradually remove tokens less relevant to the object or background patches throughout the pruning stages. This is a reasonable selection because such patches are not necessary for identifying the object being classified.

However, in audio classification tasks, it is unclear which tokens should be discarded: empty or low-intensity regions in Mel-spectrograms do not necessarily indicate a lack of information. In the task of identifying the sound of a baby crying, losing the brief silences between cries could make it more difficult to distinguish it from the sound of a siren. Furthermore, empty regions in certain frequency bands of a Mel-spectrogram can be a characteristic of the sound source, which should not be ignored.
The work most similar to ours is FastAST \cite{fastast}, which applies token merging for general audio classification.
In this work, we not only demonstrate that TopK token pruning achieves competitive accuracy on audio‐based transformer models, but also analyzes which token types contribute most to accuracy.
Our main findings are summarized as follows:
\begin{itemize}
    \item TopK token pruning based on attention scores can reduce the Multiply-Accumulate Count (MAC) operations of AudioMAE and AST by 30-40\%, with less than a 1\% drop in accuracy.
    \item By visualizing pruning patterns, we observe audio models not only retain high intensity and variation patches (bi-modal behavior) but also retain tokens from low-intensity or low-complexity Mel-spectrogram patches selectively.
    \item Based on the prior that strong, complex regions of a mel-spectrogram correspond to “object” regions (as in image tasks), we applied token pruning - prioritizing ones originated from high intensity (\textit{mean}) or variation (\textit{std}) of signals in patches.
    We found that, although pruning based on these statistics performs comparably, attention-based pruning achieves better results, suggesting that models benefit from preserving less strong and complex acoustic regions.
    \item We support this claim by measuring the high correlation (Kendall’s $\tau$) between attention scores and these statistical features, and by showing that retained tokens consistently receive distinct attention scores than pruned ones, regardless of their intensity or variation.
    \item Using selective token pruning based on intensity, we show that while high-intensity tokens contribute significantly to model performance, removing low-intensity tokens affects accuracy more in general audio classification tasks (e.g., AS-20K, ESC-50) than for speech-specific tasks (e.g., SPC-2, VoxCeleb-1).
    \item We also find that AudioMAE retains more low-intensity tokens than AST, regardless of the feature-aggregation method (using a [cls] token vs. mean pooling).
\end{itemize}

The code is publicly available at \textit{https://github.com/andylee-24/token-pruning-audio-transformer}.

\begin{figure*}[ht]
  \centering
  \includegraphics[width=17cm]{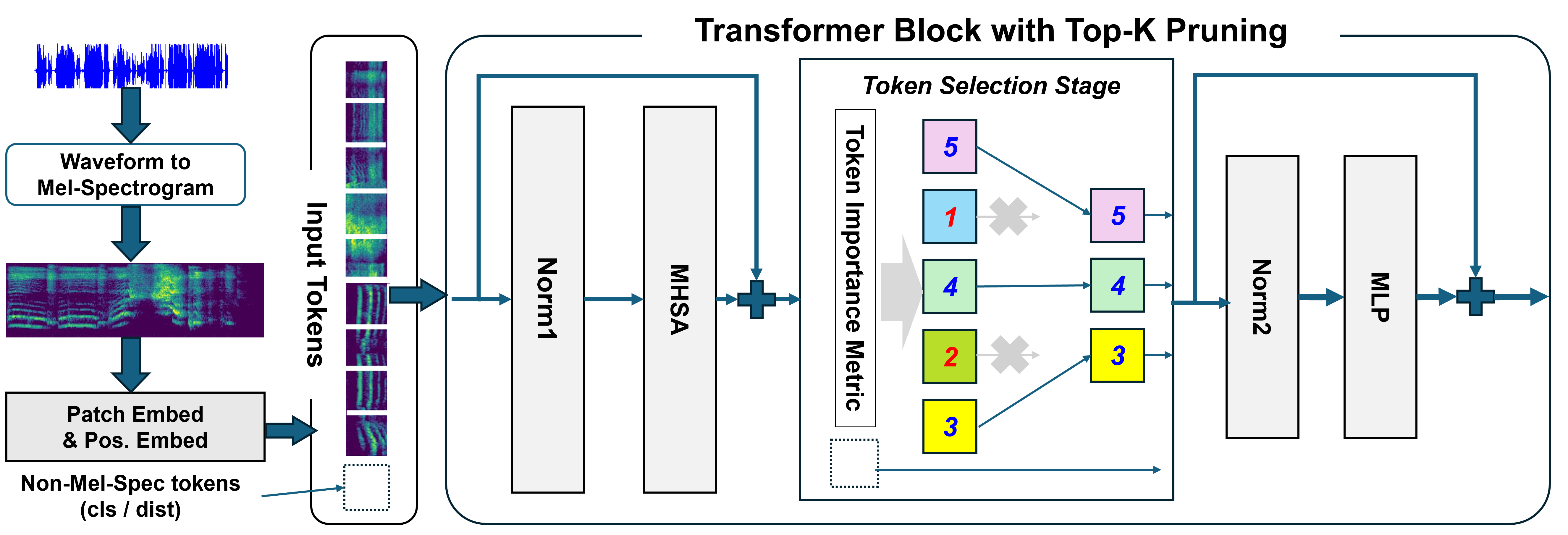}
  \caption{A transformer block equipped with TopK token pruning module. \textit{keep-rate} is set to 0.6 in this figure, so 3 out of 5 (60\%) tokens are retained.}
  \label{fig:model}
\end{figure*}

\section{Related Works}

\subsection{Audio Spectrogram Transformers}

Since the great success of the transformer architecture \citep{transformer} in Natural Language Processing (NLP), numerous works have adopted it in a variety of fields. The Vision Transformer \citep{vit} is one of the successful examples of applying the transformer to vision tasks, achieving state-of-the-art results consistently. Audio Spectrogram Transformer (AST) \citep{ast} demonstrated that the ViT is also applicable to several audio tasks by using Mel-spectrograms, which provide a visual representation of audio waveforms. Using AST, many studies have been conducted to train audio features using large amounts of unlabeled data represented as Mel-spectrograms. SS-AST \citep{ssast} explored audio feature learning through a self-supervised learning method called Masked Spectrogram Patch Modeling (MSPM). AudioMAE \citep{audiomae} extended the masked auto-encoders \citep{mae} approach to audio tasks. HTS-AT \citep{htsat} combined shifted window attention \citep{swinv1} and token semantic module for audio classification tasks. DFT-AT \citep{dftat} introduced time-frequency decoupling techniques based on MaxViT \citep{maxvit}.


\subsection{Token Reduction Methods in Vision Tasks}
\label{paper:related_works_token_reduction}

ViT-based models have achieved SOTA results across several domains, but the large computational requirements hinder their deployment. Various approaches have been studied, including efficient architecture designs \citep{mvit, swinv1}, quantization \citep{ptq_vit}, and early exit branches \citep{ee_branch_vit}. In our work, we focus on token reduction methods, which reduce the number of tokens based on the characteristics of the input data.
A-ViT \citep{avit} proposed using a single embedding dimension as a pruning score, without sub-network for pruning or extra parameters. SPViT and DynamicVit \citep{spvit, dynamicvit} determined which tokens to prune using prediction modules made of MLP layers. EViT \citep{evit} used the attention scores of tokens to determine which tokens carry more important information. ATS \citep{ats} additionally considered the norm of the value matrix in the transformer when calculating token importance scores. METR \citep{romance} showed integrating multi-exit layers with token pruning enables tokens originating from object regions to receive higher attention scores, making them more likely to be retained during the token reduction process. These methods extract attention matrix scores from the classification token, i.e. [CLS] token, as it gathers information from all tokens to form the final prediction.
Along the token pruning, token merging methods have been also studied. ToMe \citep{tome} showed that the number of tokens can be reduced through a merging strategy using the similarity between tokens without additional training. ToFu \citep{tofu} suggested applying token pruning in the earlier layers and token merging in the later layers by functional linearity analysis. LTMP ~\cite{ltmp} trains a masking module to decide which tokens to merge or prune in a ViT, taking computational constraints into account.

\subsection{Accelerating Audio Transformer Models}
DAISY and HuBERT-EE \citep{daisy, hubertee} showed that entropy can be used as an early-exit criterion for multiple branches in speech recognition tasks.
\citep{iclr25pruning} demonstrated layer-wise pruning can be applied to Whisper-style speech models conditioned on the task type.
\citep{asree} found that training Conformer models with early exit layers from scratch yields better performance than fine-tuning a pretrained model. Quantization can be applied to speech recognition models to reduce computing costs \citep{whisperquant}. FastAST \citep{fastast} combined token merging and cross-model knowledge distillation with AST model. While various techniques have been applied to accelerate audio transformers, to the best of our knowledge, no work has applied token pruning to the audio classification task and analyzed it in detail.


\begin{table*}[ht]
\caption{Benchmark results of TopK pruning on audio models; mean Average Precision (mAP) on AS-20K and Top-1 accuracy on other datasets. $\mathbf{B}$ - baseline, $kr$ - keep-rate, $\mathbf{A}$ - attention score. Section 4.2 evaluates statistics-based pruning metrics - $\mathbf{I}$: intensity ($mean$) and $\mathbf{V}$: variation ($std$).}
\label{tab:benchmark}
\centering
\setlength{\tabcolsep}{4pt} 
\resizebox{\textwidth}{!}{
\begin{tabular}{c|ccc|ccc|ccc|ccc|ccc|ccc|ccc}
\toprule
\multirow{2}{*}{-} & \multicolumn{6}{c|}{\textbf{AS-20K}} & \multicolumn{6}{c|}{\textbf{SPC-2}} & \multicolumn{6}{c|}{\textbf{ESC-50}} & \multicolumn{3}{c}{\textbf{VoxCeleb-1}} \\
\cmidrule{2-22}
 & \multicolumn{3}{c|}{\textbf{AudioMAE}} & \multicolumn{3}{c|}{\textbf{AST}} & \multicolumn{3}{c|}{\textbf{AudioMAE}} & \multicolumn{3}{c|}{\textbf{AST}} & \multicolumn{3}{c|}{\textbf{AudioMAE}} & \multicolumn{3}{c|}{\textbf{AST}} & \multicolumn{3}{c}{\textbf{AudioMAE}} \\
\midrule

$\mathbf{B}$ & \multicolumn{3}{c|}{37.8} & \multicolumn{3}{c|}{38.7} & \multicolumn{3}{c|}{98.36} & \multicolumn{3}{c|}{97.33} & \multicolumn{3}{c|}{94.10} & \multicolumn{3}{c|}{95.05} & \multicolumn{3}{c}{95.18}\\

\textit{kr} & $\mathbf{A}$ & $\mathbf{I}$ & $\mathbf{V}$ & $\mathbf{A}$ & $\mathbf{I}$ & $\mathbf{V}$ & $\mathbf{A}$ & $\mathbf{I}$ & $\mathbf{V}$ & $\mathbf{A}$ & $\mathbf{I}$ & $\mathbf{V}$ & $\mathbf{A}$ & $\mathbf{I}$ & $\mathbf{V}$ & $\mathbf{A}$ & $\mathbf{I}$ & $\mathbf{V}$ & $\mathbf{A}$ & $\mathbf{I}$ & $\mathbf{V}$ \\

0.9 & 37.3 & 36.7 & 37.0 & 38.7 & 36.6 & 36.8 & 98.09 & 98.27 & 98.33 & 97.21 & 97.08 & 97.06 & 93.66 & 93.39 & 93.75 & 94.36 & 92.05 & 92.60 & 95.05 & 95.16 & 95.21 \\
0.8 & 36.8 & 35.7 & 36.5 & 37.9 & 35.9 & 36.2 & 97.77 & 98.26 & 98.38 & 97.22 & 97.11 & 97.14 & 93.45 & 92.97 & 93.67 & 94.32 & 91.80 & 92.32 & 94.77 & 94.66 & 95.27 \\
0.7 & 36.2 & 34.5 & 35.5 & 37.8 & 35.1 & 35.7 & 97.70 & 98.18 & 98.24 & 97.19 & 96.77 & 96.98 & 93.46 & 92.25 & 93.16 & 94.37 & 91.03 & 92.13 & 94.46 & 93.37 & 94.64 \\
0.6 & 35.5 & 33.0 & 34.1 & 37.6 & 34.4 & 34.9 & 97.66 & 97.95 & 98.20 & 97.20 & 96.82 & 96.81 & 93.16 & 91.79 & 92.40 & 94.37 & 90.31 & 91.48 & 93.38 & 90.47 & 93.06 \\
0.5 & 34.4 & 31.0 & 32.4 & 37.2 & 33.4 & 34.0 & 97.44 & 97.68 & 97.85 & 97.11 & 96.72 & 96.74 & 92.75 & 90.79 & 91.23 & 94.07 & 89.36 & 89.54 & 91.26 & 86.58 & 89.54 \\
0.4 & 32.8 & 28.4 & 30.2 & 36.8 & 32.1 & 32.7 & 97.35 & 97.48 & 97.35 & 97.07 & 96.60 & 96.84 & 91.87 & 89.53 & 90.02 & 93.87 & 87.69 & 88.37 & 88.02 & 80.81 & 84.71 \\
\bottomrule
\end{tabular}
}
\end{table*}
\begin{table}[ht]
\centering
\caption{MAC(G) values across different datasets and \textit{keep-rate} from 0.4 to 1.0. $N$ denotes the number of tokens from Mel-spectrogram.}
\label{tab:mac_g_values}
\resizebox{\linewidth}{!}{
\begin{tabular}{c|r|rrrrrrr}
\toprule
\textbf{Dataset} & \textbf{$N$} & \textbf{1.0} & \textbf{0.9} & \textbf{0.8} & \textbf{0.7} & \textbf{0.6} & \textbf{0.5} & \textbf{0.4} \\
\midrule
SPC-2      &    64  & 5.6 &   4.9 &   4.30 &   3.7 &   3.3 &   2.8 &   2.5 \\
ESC-50     &  256   & 23.1 &  20.0 &  17.3 &  15.0 &  13.1 &  11.4 &  10.0 \\
\multirow{2}{*}{\shortstack{AS-20K \\ VoxCeleb-1}} & \multirow{2}{*}{512} & \multirow{2}{*}{48.6} & \multirow{2}{*}{41.8} & \multirow{2}{*}{36.0} & \multirow{2}{*}{31.1} & \multirow{2}{*}{27.1} & \multirow{2}{*}{23.7} & \multirow{2}{*}{20.8} \\
 &  &  &  &  &  &  &  \\
\bottomrule
\end{tabular}
}
\end{table}

\section{Token Pruning on Audio Transformer Models}

A Transformer processes input sequence of N tokens into an activation matrix of shape (N, D), then iteratively refines these features over multiple blocks. The Multi-Head Self-Attention (MHSA) is a key component of the Transformer. Tokens are transformed into three matrices: Q (Query), K (Key), and V (Value) through linear projections. The attention matrix $\mathbf{A}$ in (\ref{eqn:attention}) allows the model to focus on more relevant tokens for each position.

\begin{equation} \label{eqn:attention}
    \text{Attention}(\mathbf{Q}, \mathbf{K}, \mathbf{V}) = \text{Softmax}\left(\frac{\mathbf{Q}\mathbf{K}^\top}{\sqrt{d}}\right)\mathbf{V}=\mathbf{A}\mathbf{V}
\end{equation}

\begin{subequations}
    \begin{align}
        a_{i, \text{mean-pooling}} &= \frac{1}{HN}\sum_{h=1}^{H}\sum_{n=1}^{N}\mathbf{A}[h,n,i] \label{eqn:mp_topk} \\
        a_{i,\text{cls}} &=\frac{1}{H}\sum_{h=1}^{H}\mathbf{A}[h,0,i] \label{eqn:cls_topk}
    \end{align}
\end{subequations}

\subsection{TopK Token Pruning on Audio Classification Transformers}

To apply token pruning, we adopted AudioMAE and AST as they are representative audio classification models trained by masked auto-encoding and supervised training, respectively. We use TopK as a token pruning method since it is a competitive method~\cite{whichtokentouse} among other SOTA methods. It also allows us to distinguish the origin of tokens for the analysis in the following section. After the raw waveform is converted into a Mel-spectrogram, it is treated as an image so that patch embeddings and positional embeddings can be applied. The token pruning module is placed between the multi-head self-attention module and the MLP module of selected (the 4th, 7th, and 10th) ViT blocks as in Fig.~\ref{fig:model}. The choice of pruning location follows the previous token pruning works~\cite{ats, evit, romance, dynamicvit}. In AudioMAE, we used token-to-token attention scores in (\ref{eqn:mp_topk}), which quantify the attention each token receives from others as shown in (\ref{eqn:attention}) - as indicators of token importance since mean pooling is used for the final prediction. $H$ and $N$ indicate the number of heads and the number of tokens from Mel-spectrogram, respectively. In AST, we used CLS attention scores (\ref{eqn:cls_topk}), following~\cite{evit}. Among the $N$ tokens, we retained $(N \times \textit{keep-rate})$ tokens with the highest attention scores in each pruning block.
The same \textit{keep-rate} is applied to all pruning-enabled blocks.

\subsection{Datasets and Metrics}
We trained and evaluated models on widely adopted audio classification datasets: AudioSet Balanced (AS-20K) ~\cite{audioset}, Speech Commands V2 (SPC-2) \cite{spc2}, Environmental Sound Classification (ESC-50) \cite{esc50} and VoxCeleb-1 \cite{voxceleb}. The number of classes is 527, 35, 50 and 1251 respectively. 
We reported the maximum mAP for AS-20K and Top-1 accuracy for other datasets.
Since AST shows low accuracy on VoxCeleb-1 (41.1\%)~\cite{ssast}, we excluded it from our experiments.

\subsection{Training Hyperparameters}
We downloaded checkpoints of AudioMAE and AST pre-trained on AudioSet-2M, using a ViT-B configuration with $(16, 16)$-sized tokens without strides. For \textit{keep-rate} scheduling we adopted the method used in EViT~\cite{evit}. We followed the original training procedures and outline the modified hyperparameters in the order of AS-20K, SPC-2, ESC-50, VoxCeleb-1. 
For AudioMAE, batch sizes are 16, 256, 64, 32 and the minimum learning rate is set to $10^{-5}$ for all datasets. We trained AudioMAE for 60, 90, 120, 90 epochs for four benchmarks, with \textit{keep-rate} reduction starting at the 30th, 10th, 20th, 20th epoch and continuing for 20, 30, 40, 40 epochs. Before token pruning starts, we applied masking ratio 0.3 for AS-20K and ESC-50; 0.0 for others. Once pruning is enabled, all masking strategies including SpecAug~\cite{park2019specaugment} were disabled, since pruning itself already applies strong masking. For AST, batch sizes were set to 64, 128, 48 and learning rates to $10^{-4}$, $2.5\times10^{-4}$, $10^{-5}$. We trained AST for 30 epochs for all dataset, enabling pruning at the 15th, 5th, 5th epoch and reducing \textit{keep-rate} for 10, 15, 15 epochs.
We used PyTorch~\cite{pytorch} with two GPUs and automatic mixed precision \cite{amp} for our training, and torchprofile~\cite{torchprofile} to measure MACs.

\begin{figure*}[ht]
  \centering
  \includegraphics[width=\linewidth]{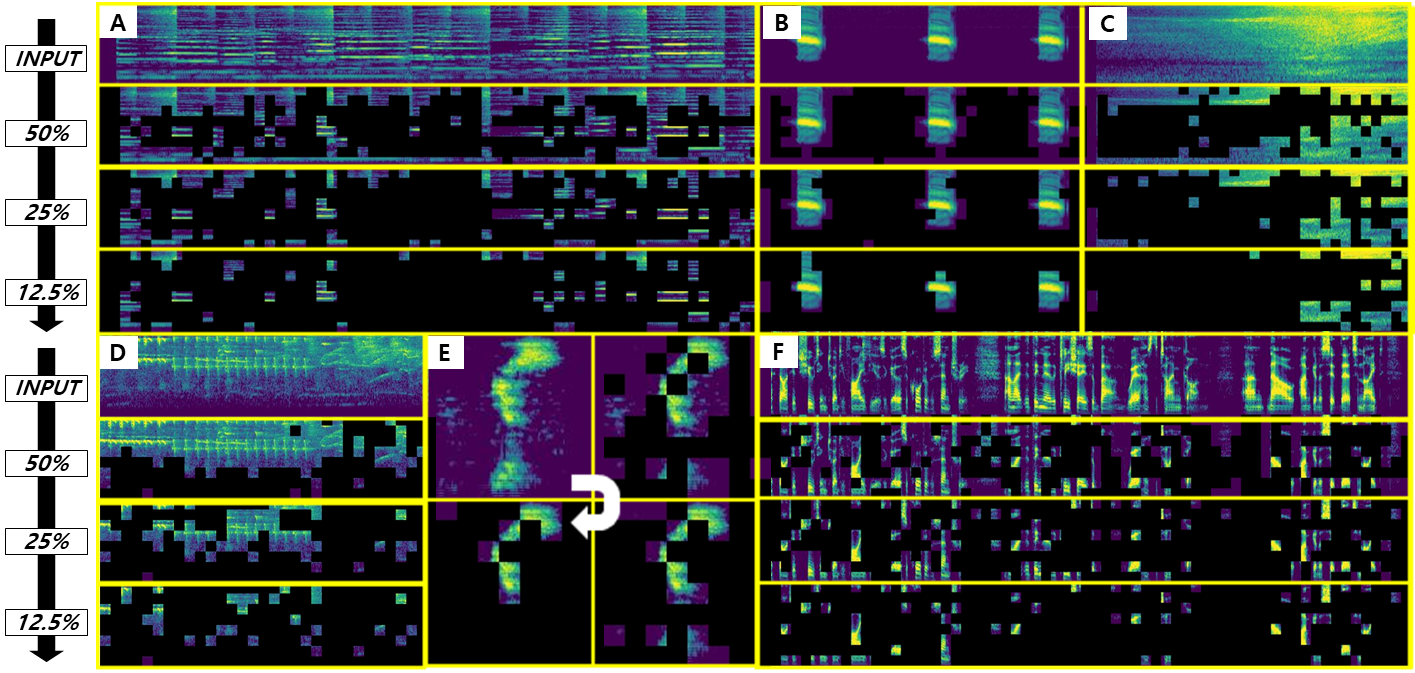}
  \caption{Visualization of TopK token pruning patterns on audio models. \textit{AST}: A / AS-20K / Brass-instrument, B / ESC-50 / Frog. \textit{AudioMAE}: C / ESC-50 / Airplane, \textit{AudioMAE}: D / ESC-50 / Birds, E / SPC-2 / a keyword "yes",  F / VoxCeleb-1 / voice. In all examples, the \textit{keep-rate} is set to 0.5.}
  \label{fig:audiomae_visualization}
\end{figure*}

\subsection{Benchmark Results of TopK Token Pruning on Audio Models}\label{sec:tables}

Table~\ref{tab:benchmark} and~\ref{tab:mac_g_values} demonstrate that a simple TopK pruning based on attention scores can reduce the computation (MAC) by 30-40\%, with less than a 1\% drop in accuracy with both audio transformer models. We also found that AudioMAE is more sensitive to token loss than AST especially at lower keep‑rates, likely due to its reliance on multiple tokens for the final prediction. Furthermore, accuracy tends to drop more when more tokens are pruned—particularly for harder tasks with a larger number of classes.
AudioMAE’s performance degrades more than AST’s as the keep‑rate decreases. We attribute this to the reduced number of tokens used in mean pooling for classification. 
We compare AST with TopK pruning against FastAST~\cite{fastast}, which adopts token merging (ToMe) ~\cite{tome} on our benchmarks.
Since FastAST uses overlapping patch embedding layers ($stride=10$), its number of input tokens differs from ours.
Therefore, for each merging ratio $r$—the number of tokens merged at each block—we compare the accuracy at the keep‑rate whose MAC reduction ratio is closest to that of the baseline ($r=0$, $kr=1.0$).
Table~\ref{tab:fastast_compare} shows that attention‑based TopK token pruning performs comparably to the token‑merging method (ToMe).
\begin{table}[htbp]
\centering
\caption{Comparison between FastAST (ToMe) and TopK pruning.}
\label{tab:fastast_compare}
\resizebox{\linewidth}{!}{

\begin{tabular}{c|r r r|r r r}
\toprule
\multirow{2}{*}{\textbf{Dataset}} & \multicolumn{3}{c|}{\textbf{ToMe (FastAST, train+inf)}} & \multicolumn{3}{c}{\textbf{TopK (Ours)}} \\
\cmidrule(lr){2-7}
 & r & \(\Delta MAC\) & acc / drop & \textit{kr} & \(\Delta MAC\) & acc / drop\\
\midrule
\multirow{3}{*}{AS-20K} 
  & 0   & 100\% & 38.2    & 1   & 100\% & 38.7    \\
  & 20 & 88.4\%  & 37.6 (-0.6) & 0.9 & 86.0\%  & 38.7 (-0.0) \\
  & 40 & 77.1\%  & 36.9 (-1.3) & 0.8 & 74.1\%  & 37.9 (-0.8) \\
\midrule
\multirow{3}{*}{ESC-50} 
  & 0   & 100\% & 94.7    & 1   & 100\% & 95.05    \\
  & 20 & 78.6\%  & 94.3 (-0.4) & 0.8 & 74.9\%  & 94.32 (-0.73) \\
  & 40 & 57.1\%  & 92.9 (-1.8) & 0.6 & 56.7\%  & 94.37 (-0.68) \\
\midrule
\multirow{3}{*}{SPC-2} 
  & 0   & 100\% & 97.8    & 1   & 100\% & 97.33    \\
  & 5  & 71.5\%  & 97.7 (-0.1) & 0.7 & 66.1\%  & 97.19 (-0.14) \\
  & 10 & 44.6\%  & 97.5 (-0.3) & 0.4 & 44.6\%  & 97.07 (-0.26) \\
\bottomrule
\end{tabular}
}
\end{table}

\section{Analysis of TopK Pruning on Audio Models}

\subsection{Visualization of Token Pruning Patterns on Audio Models}
We begin our analysis of important tokens by visualizing TopK token pruning patterns of AudioMAE and AST in Fig.~\ref{fig:audiomae_visualization}. The models effectively discard tokens originating not only from the padding regions (\textbf{A, E, F}) but also from pauses between sounds (\textbf{B}).
This observation suggests that models preferentially retain tokens from high-intensity Mel-spectrogram patches.
METR~\cite{romance} showed that the ViT‑based image classifiers perform better when the token‑pruning module retains tokens from object regions than background regions. In Mel-spectrograms of the SPC-2 dataset, we can easily distinguish between tokens representing signals and those representing noise/background, as each sample contains only a single word.
However, in Fig.~\ref{fig:audiomae_visualization} (E), AudioMAE discarded some tokens belonging to a word.
We can find similar pruning pattern appears in the first pruning stage of VoxCeleb-1 in (\textbf{F}), tokens originated from high-intensity patches are discarded instead of low intensity tokens.
In summary, some tokens from low-intensity regions are retained even if they are not part of padding or pause regions - indicating that they received lower attention scores than certain low-intensity patches.
\subsection{Token Pruning with Statistical Features}

Based on this observation, we wondered whether we can prune tokens using features directly available from the Mel-spectrogram. Specifically, we considered the intensity ($mean$ of signals in a $16\times 16$ patch), as high-intensity patches typically carry more energy, and variation ($std$ of signals in a $16 \times 16$ patch), as it can capture the patch’s texture complexity in the Mel-spectrogram. This motivated us to explore whether these statistics could serve as alternative pruning metrics to attention scores. We fine-tuned both models using $mean$ and $std$ as pruning metrics, replacing attention scores while keeping the same hyperparameters. In Table~\ref{tab:benchmark}, these statistics perform as well as attention scores at high \textit{keep-rates} (e.g., 0.7), but fail to outperform attention scores when tokens are aggressively pruned. SPC-2 shows minimal impact because the input data only contains a spoken keyword.
This suggests that audio models also needs to retain tokens originated from low-intensity or low-variation patches for better prediction.

\subsection{Correlation Between Attention and Statistical Features}

\begin{figure}[ht]
  \centering
  \includegraphics[width=\linewidth]{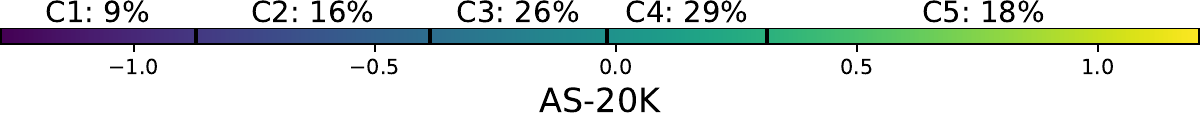}
  \includegraphics[width=\linewidth]{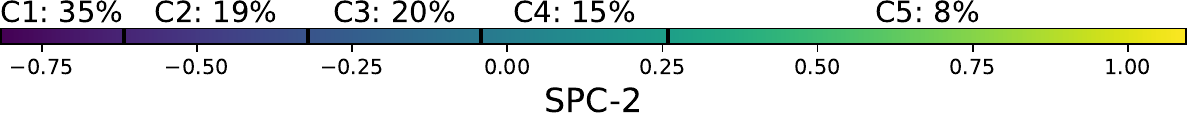}
  \includegraphics[width=\linewidth]{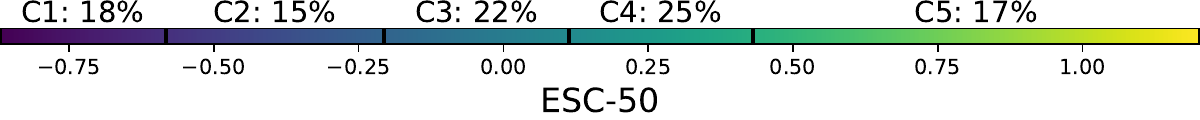}
  \includegraphics[width=\linewidth]{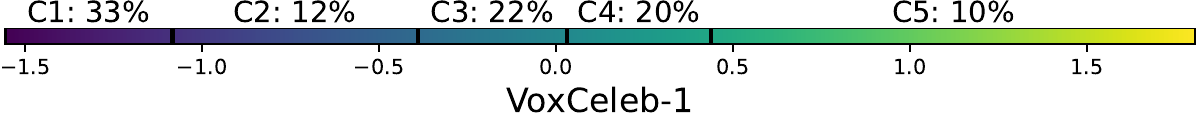}
  \caption{Ranges of clustered tokens’ normalized intensities obtained via K‑means clustering for each dataset. The horizontal color bar shows the normalized mel‑spectrogram's intensity scale. Percentages next to each cluster name denote the proportion of tokens in the cluster.}
  \label{fig:cluster_mean}
\end{figure}

As shown in \cite{romance}, in image classification tasks ViT performs better if tokens corresponding to objects are not pruned. This is reasonable because clear boundaries exist between objects and the background in images. However, it is hard to say that audio models prunes tokens well merely by observing the pattern since low-intensity regions might still contribute to class discrimination. To investigate the models' preference over high-intensity regions, we quantify the preference using Kendall's $\tau$ correlation~\cite{kendall} between intensity $(mean)$ and variation $(std)$ of each patch and the corresponding attention score. Clustering is necessary to avoid ranking reordering due to minor differences in intensity (or variation), which is not perceptually differentiable. We used K-means clustering with size 5 for both intensity (Fig.~\ref{fig:cluster_mean}) and variation (Table~\ref{tab:cluster_std}).

For token index $i$, $m(i)$ and $std(i)$ denote the mean normalized intensity and standard deviation of token's i's corresponding mel-spectrogram patch. We clustered $m(i)$ into 5 groups using K-means, collecting tokens from the evaluation dataset for each task. We denote $C(m(i))$ as token i's corresponding cluster index sorted in increasing mean intensity and $a_i$ as an attention score.
A token pair is concordant if satisfies (\ref{eqn:concordant_1}) or (\ref{eqn:concordant_2}), and discordant otherwise.

\begin{subequations}
\begin{align}
    (C(m(i)) \leq C(m(j)) \land (a_i \leq a_j) \label{eqn:concordant_1} \\ 
    (C(m(i)) > C(m(j)) \land (a_i > a_j) \label{eqn:concordant_2}
\end{align}
\end{subequations}

\begin{equation} \label{eqn:kendall}
    \tau=\frac{\text{\# concordant pairs} - \text{\# discordant pairs}}{\text{\# pairs}}
\end{equation}

\begin{figure}[ht]
  \centering
  \includegraphics[width=\linewidth]{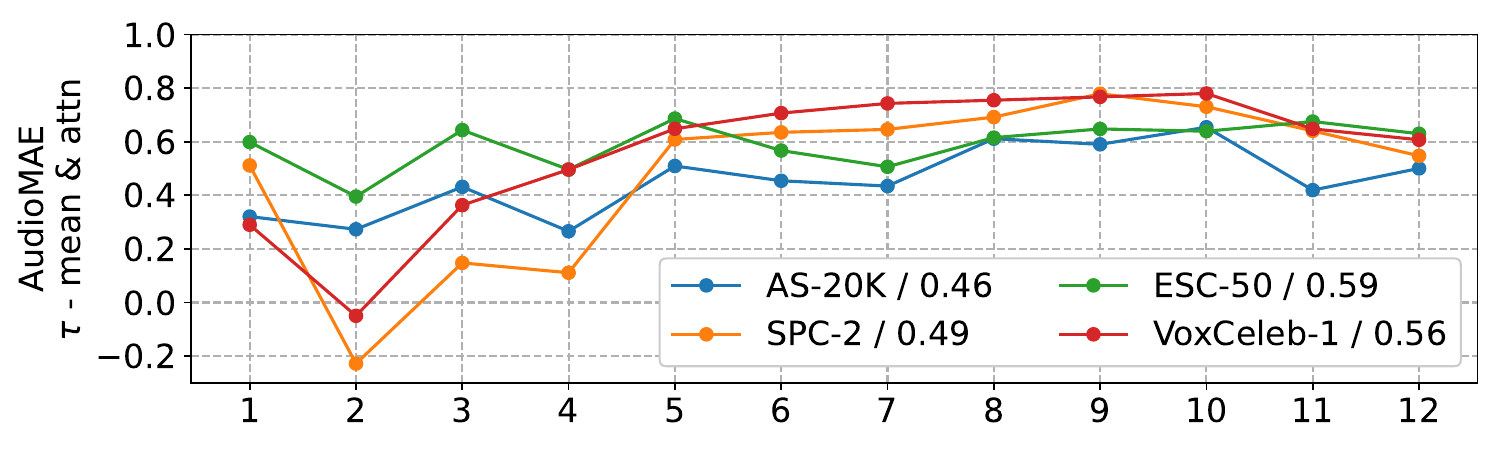}
  \includegraphics[width=\linewidth]{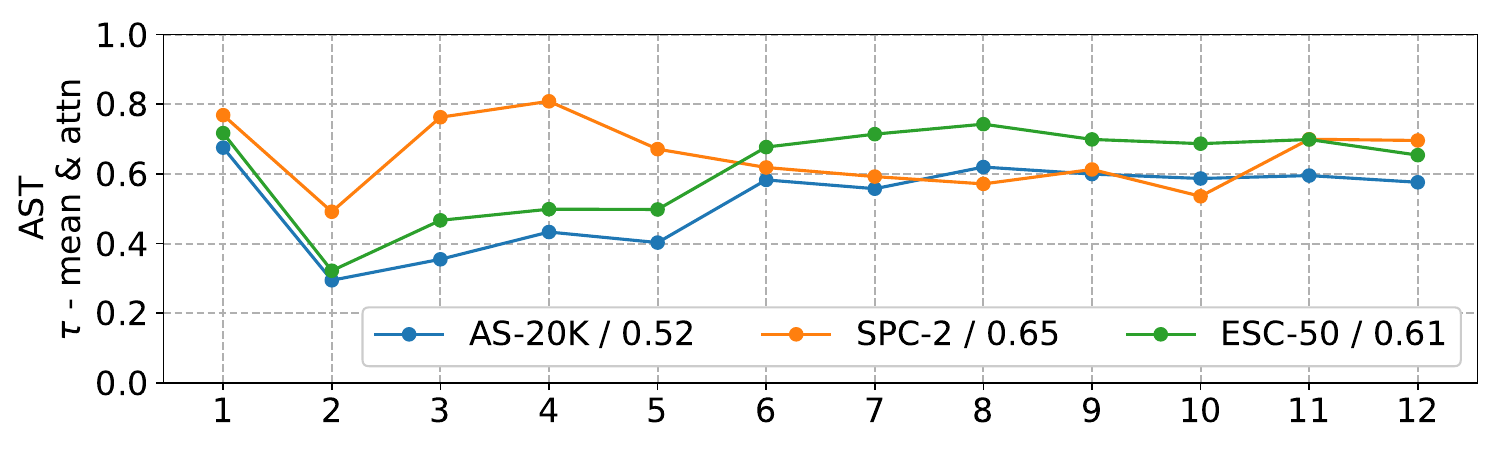}
  \includegraphics[width=\linewidth]{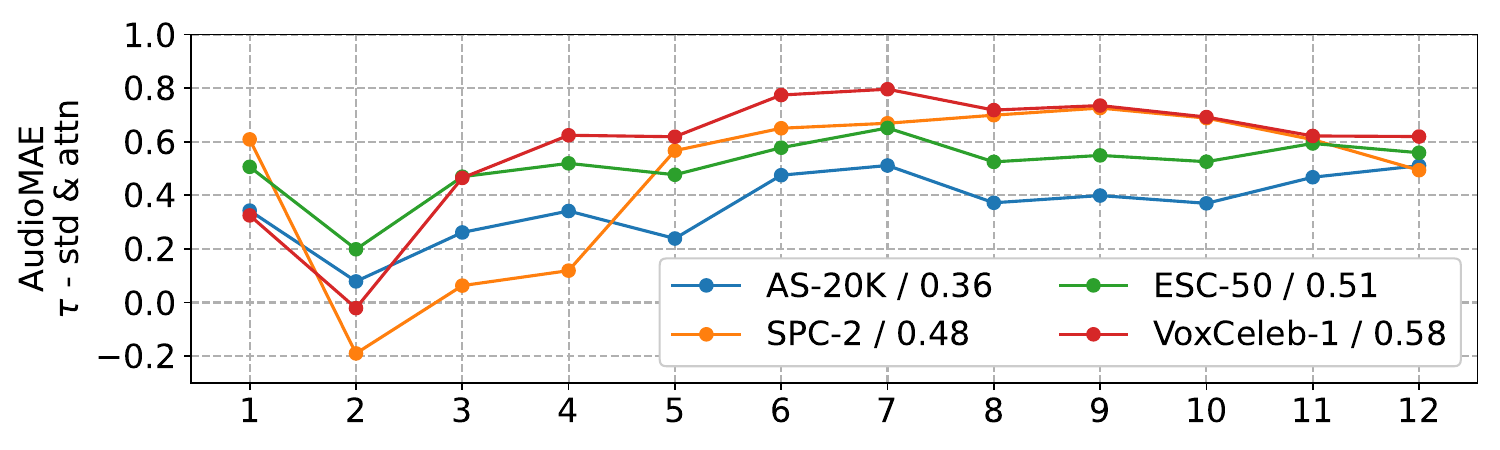}
  \includegraphics[width=\linewidth]{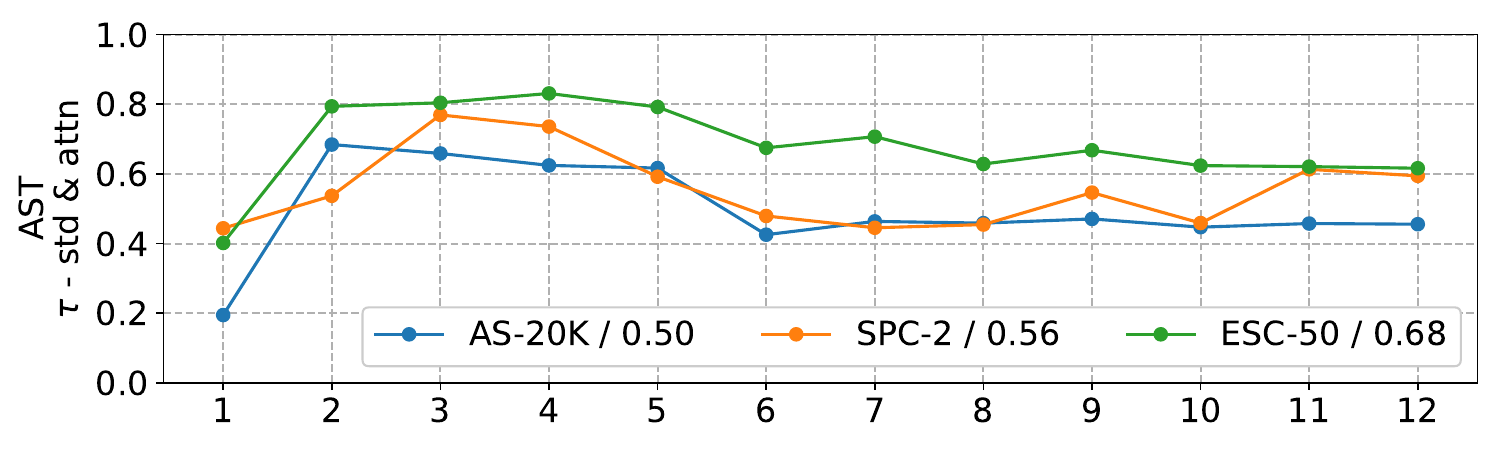}
  \caption{Kendall's $\tau$ correlation between token's normalized intensity $(mean)$ / variation $(std)$ and attention score. The numbers following each dataset name indicates the average $\tau$ across 12 blocks.}
  
  \label{fig:kendall_rank_intensity}
\end{figure}

Similarly, we use the same condition by replacing $m()$ by $std()$. The Kendall $\tau$ is given by (\ref{eqn:kendall}).
Fig.~\ref{fig:kendall_rank_intensity} shows that both statistics are positively correlated with attention score, which explains the comparable performance to the attention based TopK model.

\begin{table}[ht]
\centering
\caption{Cluster sizes obtained by K-means with standard deviation.}
\label{tab:cluster_std}
\resizebox{6.5cm}{!}{
\begin{tabular}{cccccc}
\toprule
\textbf{Dataset} & \textbf{C1} & \textbf{C2} & \textbf{C3} & \textbf{C4} & \textbf{C5} \\
\midrule
\textbf{AS-20K}    & 23.1       & 43.2       & 20.4       & 11.9       & 1.5        \\
\textbf{SPC-2}      & 18.5       & 33.8       & 27.5       & 15.0       & 5.2        \\
\textbf{ESC-50}     & 12.1       & 47.5       & 23.1       & 13.4       & 3.9        \\
\textbf{VoxCeleb-1}& 32.9       & 23.5       & 24.7       & 14.1       & 4.8        \\
\bottomrule
\end{tabular}
}
\end{table}

\begin{table*}[ht]
\caption{Attention score ratio over groups ($\Gamma$) on different datasets, models and \textit{keep-rate}.}
\label{tab:Gamma}
\centering
\setlength{\tabcolsep}{4pt} 
\resizebox{\textwidth}{!}{
\begin{tabular}{c|rrr|rrr|rrr|rrr|rrr|rrr|rrr}
\toprule
\multirow{2}{*}{-}            
  & \multicolumn{6}{c|}{\textbf{AS‑20K}}
  & \multicolumn{6}{c|}{\textbf{SPC‑2}}
  & \multicolumn{6}{c|}{\textbf{ESC‑50}}
  & \multicolumn{3}{c}{\textbf{VoxCeleb‑1}} \\ \cmidrule(lr){2-22}
  & \multicolumn{3}{c|}{\textbf{AudioMAE}}
  & \multicolumn{3}{c|}{\textbf{AST}}
  & \multicolumn{3}{c|}{\textbf{AudioMAE}}
  & \multicolumn{3}{c|}{\textbf{AST}}
  & \multicolumn{3}{c|}{\textbf{AudioMAE}}
  & \multicolumn{3}{c|}{\textbf{AST}}
  & \multicolumn{3}{c}{\textbf{AudioMAE}} \\ \midrule
\textit{kr}
  & $\Gamma_{1}$ & $\Gamma_{2}$ & $\Gamma_{3}$ & $\Gamma_{1}$ & $\Gamma_{2}$ & $\Gamma_{3}$
  & $\Gamma_{1}$ & $\Gamma_{2}$ & $\Gamma_{3}$ & $\Gamma_{1}$ & $\Gamma_{2}$ & $\Gamma_{3}$
  & $\Gamma_{1}$ & $\Gamma_{2}$ & $\Gamma_{3}$ & $\Gamma_{1}$ & $\Gamma_{2}$ & $\Gamma_{3}$
  & $\Gamma_{1}$ & $\Gamma_{2}$ & $\Gamma_{3}$ \\ \midrule
0.8
  & 1.15 & 1.61 & 3.00 & 1.29 & 3.57 & 4.51
  & 1.27 & 2.89 & 7.02 & 5.85 & 3.52 & 3.63
  & 1.15 & 1.75 & 5.52 & 2.01 & 5.09 & 31.37
  & 1.24 & 6.52 & 17.52 \\
0.6
  & 1.14 & 1.47 & 2.22 & 1.27 & 2.91 & 3.87
  & 1.18 & 2.03 & 2.93 & 4.96 & 2.78 & 2.79
  & 1.13 & 1.54 & 2.19 & 2.01 & 4.02 & 12.49
  & 1.22 & 3.22 & 2.91 \\
0.4
  & 1.18 & 1.41 & 1.69 & 1.29 & 2.52 & 2.94
  & 1.13 & 1.43 & 1.61 & 4.03 & 2.33 & 2.31
  & 1.12 & 1.45 & 1.54 & 2.06 & 3.24 & 4.69
  & 1.20 & 1.50 & 1.91 \\ \bottomrule
\end{tabular}
}
\end{table*}

\begin{figure*}[ht]
  \centering
\resizebox{\linewidth}{!}{
  \begin{tabular}{ccc}
  \includegraphics[width=\linewidth]{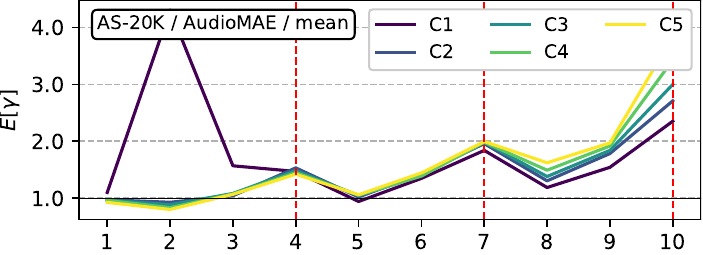} & 
  \includegraphics[width=\linewidth]{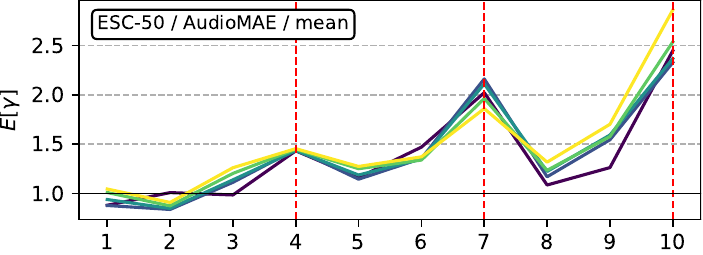} &
  \includegraphics[width=\linewidth]{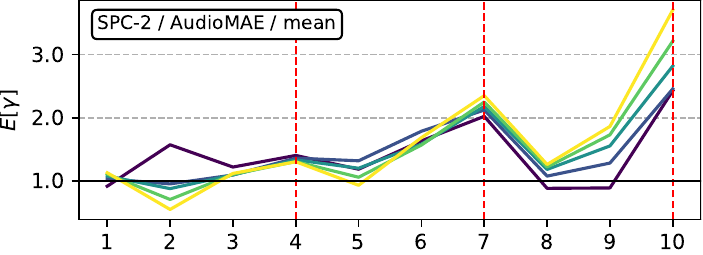} \\
  \includegraphics[width=\linewidth]{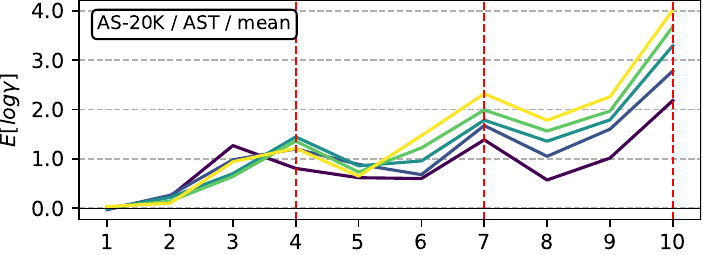}  & 
  \includegraphics[width=\linewidth]{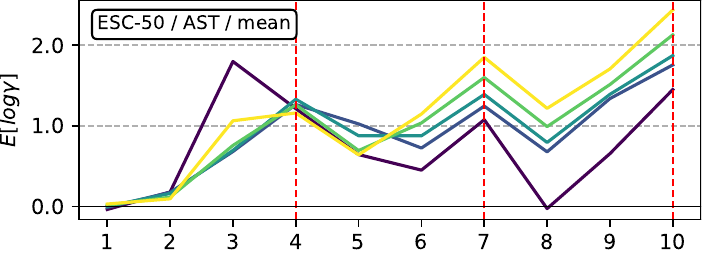} & 
  \includegraphics[width=\linewidth]{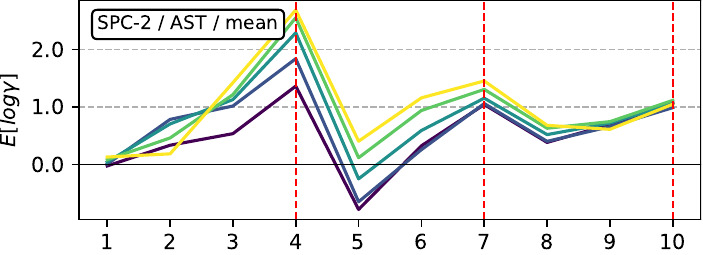} \\
\includegraphics[width=\linewidth]{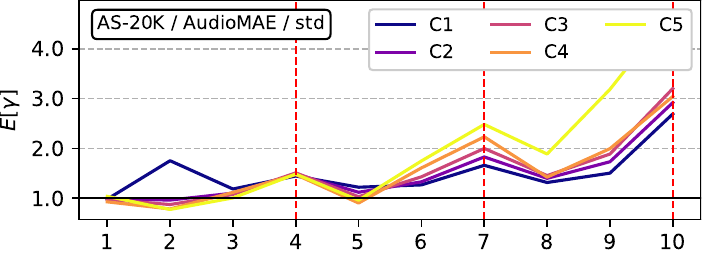} & 
  \includegraphics[width=\linewidth]{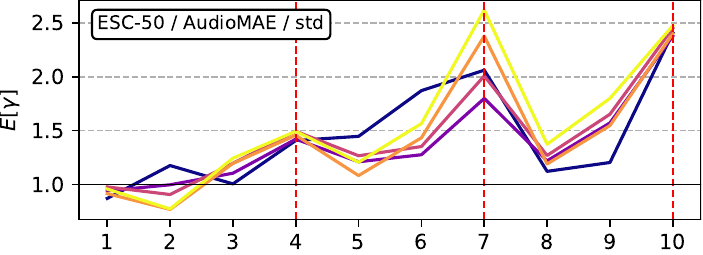} &
  \includegraphics[width=\linewidth]{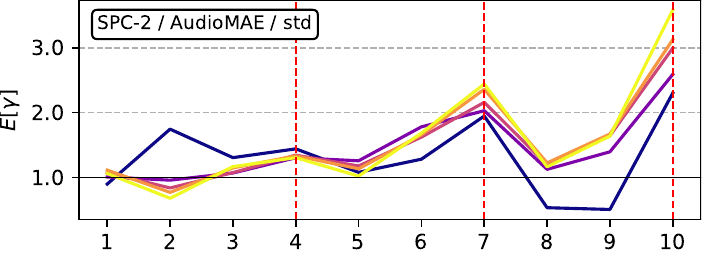} \\
  \includegraphics[width=\linewidth]{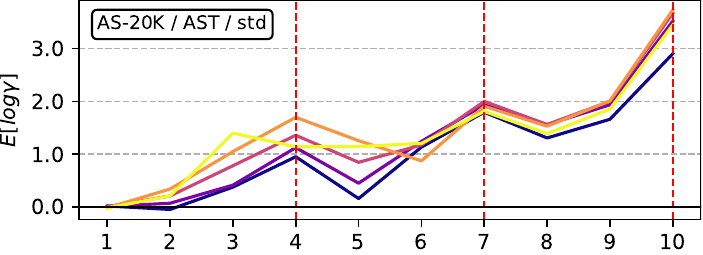}  & 
  \includegraphics[width=\linewidth]{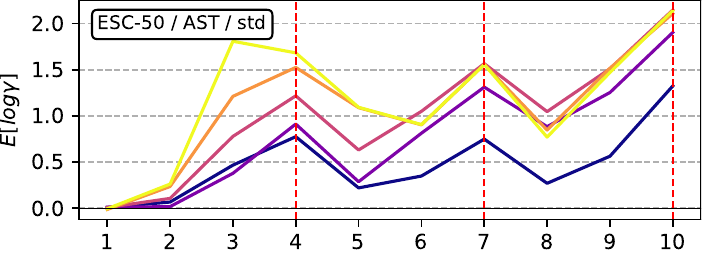} & 
  \includegraphics[width=\linewidth]{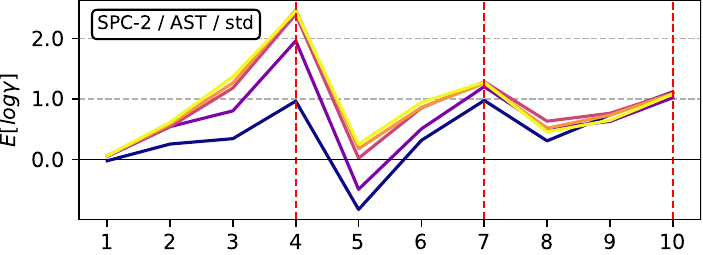} 
    \end{tabular}
  }
    \caption{Attention score ratio $\gamma$ across models and datasets. Since AST shows larger $\gamma$ near pruning, we plot $\mathbb E[\log\gamma]$ for AST and $\mathbb E[\gamma]$ for AudioMAE over blocks for others to illustrate changes of attention scores clearly. \textit{keep-rate} is set to 0.5. Red lines indicates at the 4th, 7th, and 10th blocks.}
  \label{fig:attention_score_ratio}
\end{figure*}

\subsection{Comparing the Attention Score of Retained Tokens and Pruned Tokens}
To support our claim that low‑intensity/variation tokens are important, we measured how confidently the model selects tokens to retain or prune.
We group attention scores by three pruning locations: $G_1=(1,2,3,\textbf{4})$ / $G_2=(5,6,\textbf{7})$ / $G_3=(8,9,\textbf{10})$ and split tokens in each group into retained and pruned sets.
For block $b$, $R_i^b$ and $P^b$ denote the sets of retained tokens in cluster $C_i$ (\textit{mean} or \textit{std}) and tokens going to be pruned at their corresponding pruning block, respectively. (e.g., $R_i^5=R_i^6=R_i^7$ and $P^5=P^6=P^7$).
We define Equation (\ref{eqn:attn_score}) representing the average ratio of attention scores for retained tokens in cluster $C_i$ to those for pruned tokens at block $b$.
We use this ratio to track attention values of retained versus pruned tokens at each block in each cluster and Equation (\ref{eqn:attn_score_group}) averages these ratios across blocks within the same group.
\begin{subequations}
    \begin{align}
\gamma(b, i)=\frac{\mathbf{E}\left[\text{attention\_score}(R_i^b)\right]}{\mathbf{E}\left[\text{attention\_score}(P^b)\right]} \label{eqn:attn_score} \\
    \Gamma(n)=\frac{1}{5|G_n|}\sum_{i=1}^{5} \sum_{j}{\gamma(G_n[j], i)} \label{eqn:attn_score_group}
    \end{align}
\end{subequations}
Fig.~\ref{fig:attention_score_ratio} shows $\gamma$ varies across tasks, models and blocks.
Both models assign significantly higher attention scores to retained low‑intensity tokens than to discarded ones at pruning stages, even when patches have low intensity or variation. This gap grows in later pruning stages for general audio classification tasks (AS‑20K, ESC‑50) and tokens from high‑intensity or high‑variation patches tend to receive higher attention scores than those from low‑intensity or low‑variation patches.
The local maxima patterns at the first and second pruning points suggest the model is relatively uncertain right after those stages, becoming more confident as computations proceed.
In Fig.~\ref{fig:kendall_rank_intensity}, AudioMAE shows a weaker correlation between attention scores and patch intensity or variation before the first pruning stage. This is reflected in the higher attention scores for low‑intensity or low‑variation tokens compared to high‑intensity or high‑variation tokens in Fig.~\ref{fig:attention_score_ratio}. Table~\ref{tab:Gamma} details the comparison of $\Gamma$ between different keep-rates and pruning stages.
As the keep‑rate decreases, the attention score ratio also decreases across all groups.
We explain this trend as follows: more irrelevant tokens are dropped while important ones are retained, resulting in a more even distribution of attention scores.

\begin{figure*}[ht]
  \centering
  \resizebox{\linewidth}{!}{
  \begin{tabular}{cccccc}
    \includegraphics[width=0.167\linewidth]{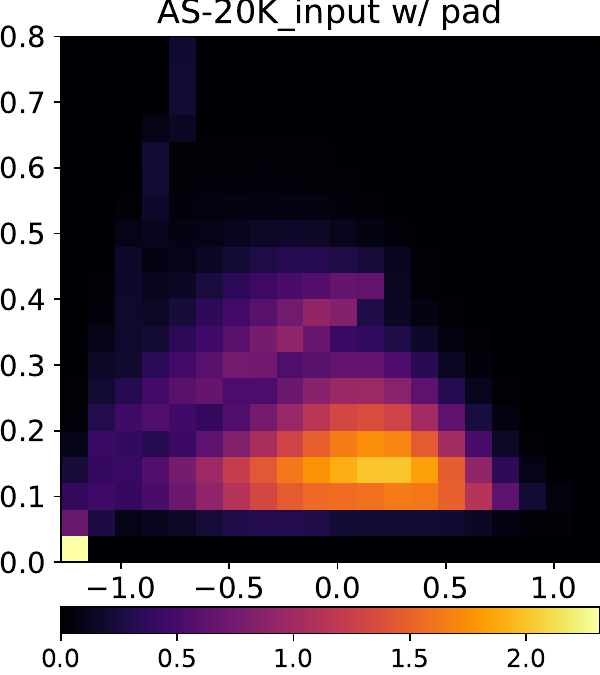} & 
    \includegraphics[width=0.167\linewidth]{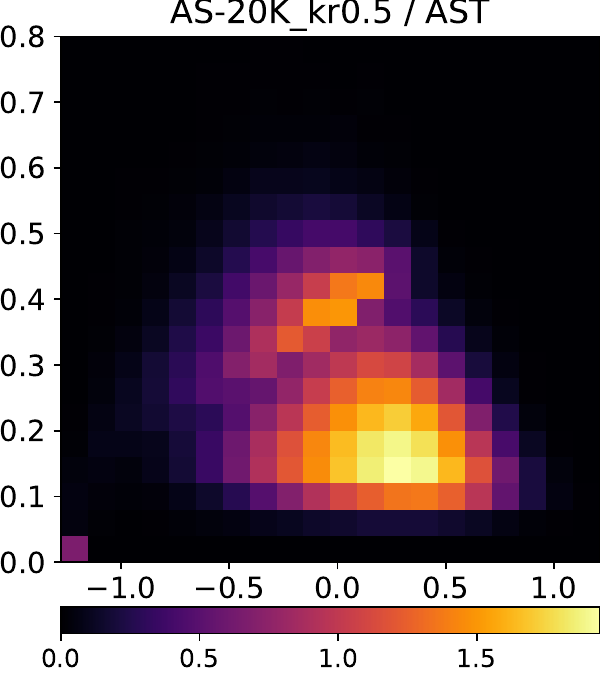} & 
    \includegraphics[width=0.167\linewidth]{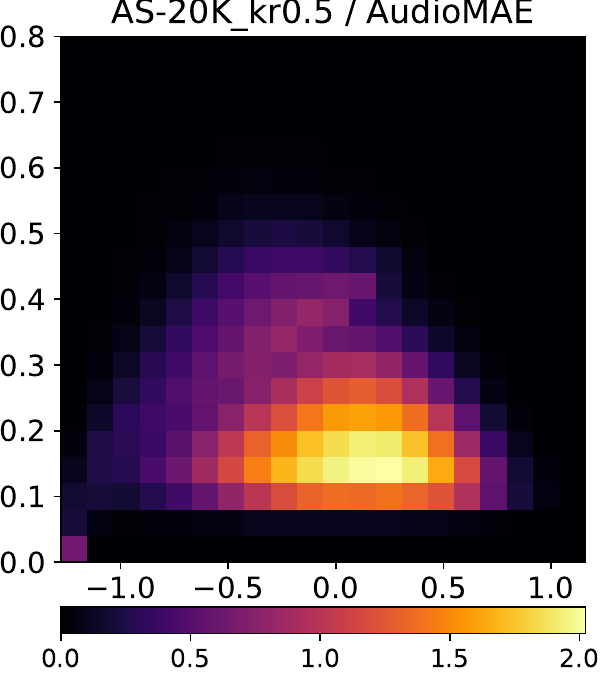} & 
    \includegraphics[width=0.167\linewidth]{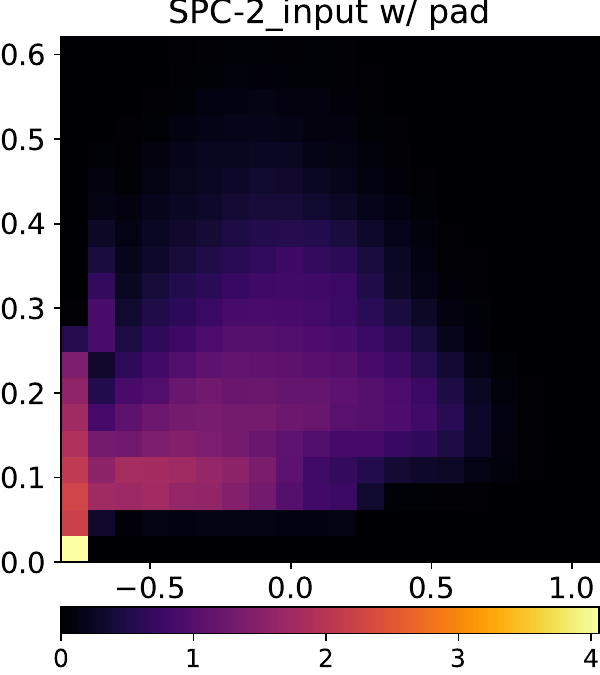} & 
    \includegraphics[width=0.167\linewidth]{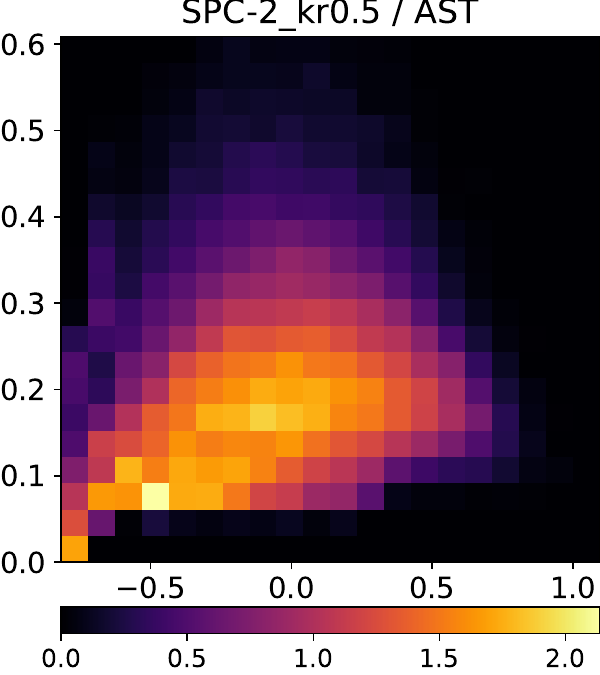} & 
    \includegraphics[width=0.167\linewidth]{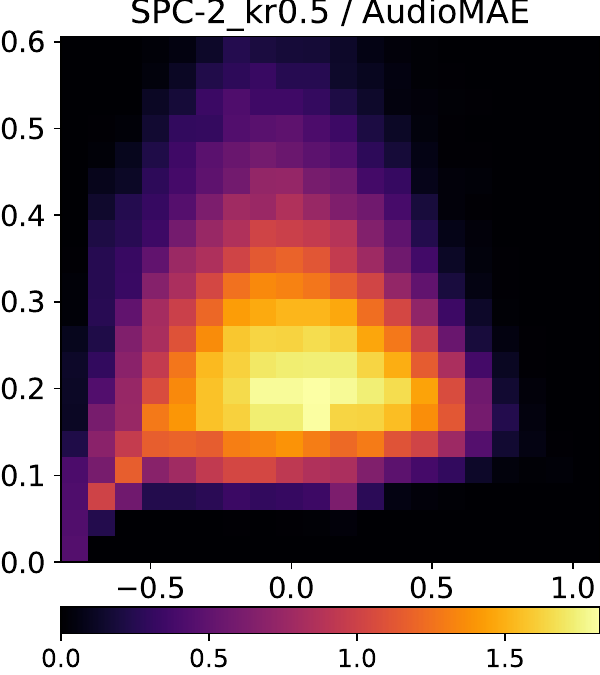} \\
    \includegraphics[width=0.167\linewidth]{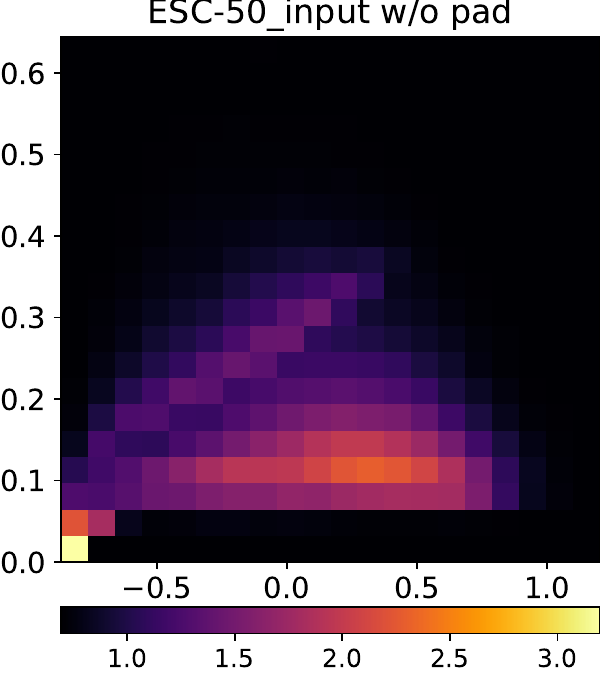} & 
    \includegraphics[width=0.167\linewidth]{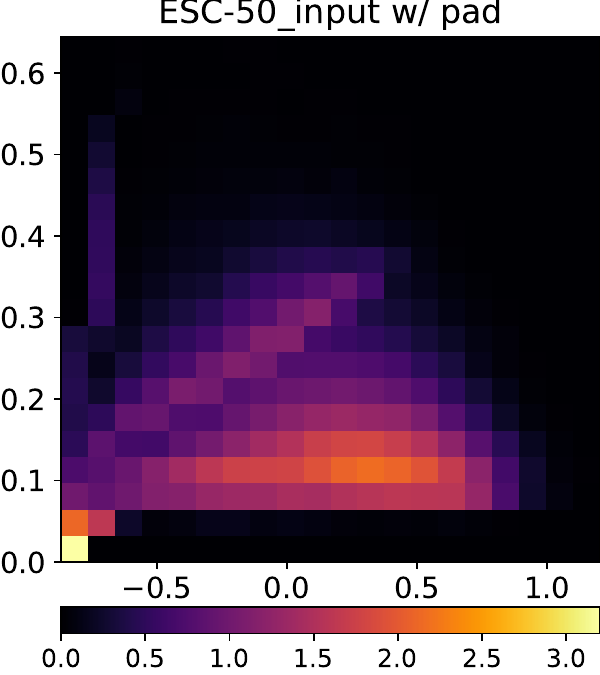} & 
    \includegraphics[width=0.167\linewidth]{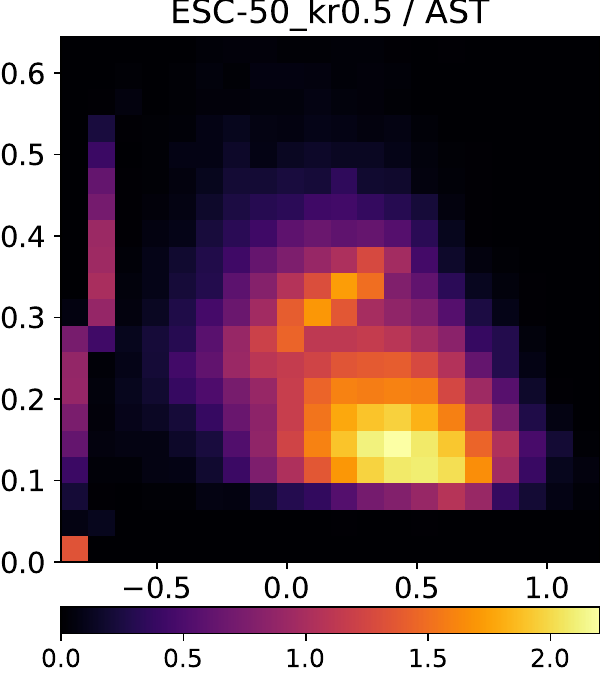} & 
    \includegraphics[width=0.167\linewidth]{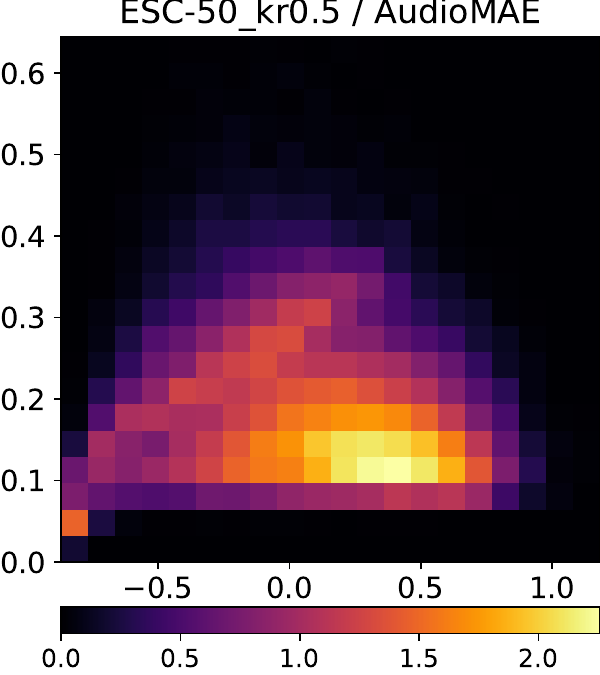} & 
    \includegraphics[width=0.167\linewidth]{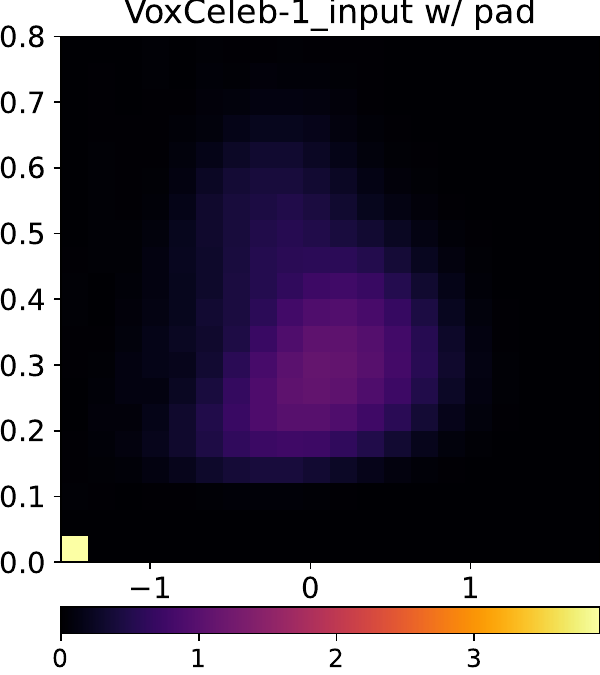} & 
    \includegraphics[width=0.167\linewidth]{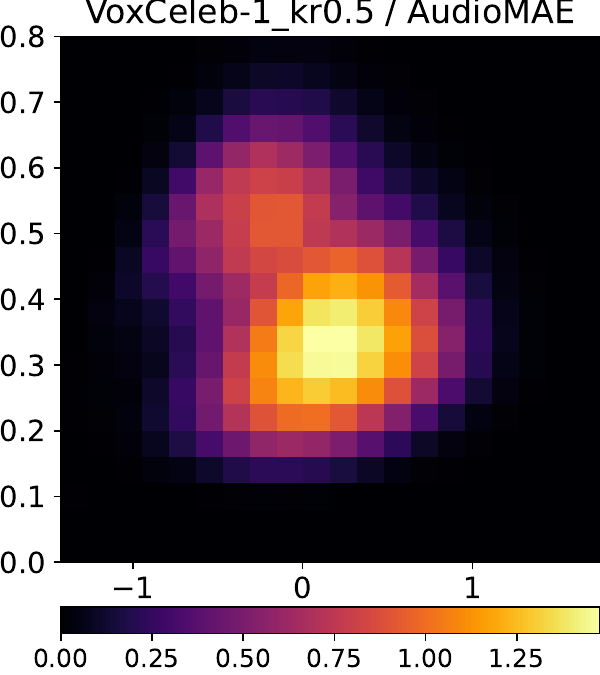} 
  \end{tabular}
  }
  \caption{Log-normalized histogram of the intensity $(mean)$ (X-axis) and variation $(std)$ (Y-axis) of signals in input patches and retained Mel-spectrogram patches of each dataset. \textit{keep-rate} is set to 0.5. In the figure for ESC-50 without (w/o) padding, we removed patches belonging to the end of the audio samples.}
  \label{fig:histogram}
\end{figure*}

\subsection{Inspection of Retained Patches' Statistics}
We visualize the relationship between the $mean$ and $std$ of signals in retained patches for each dataset in Fig.~\ref{fig:histogram}. The vertical lines on low-mean patches in the input Mel-spectrograms of ESC-50 and SPC-2 indicate artifacts from padding regions at the end of the audio samples; AudioMAE effectively discards these regions, whereas AST does not.
Except for SPC-2, which consists of a single word, there are two patch groups on either side of the diagonal in each histogram. The retention of these groups after pruning indicates the audio model's reliance on (1) low-to-mid intensity patches with greater complexity and (2) high-intensity patches with less complexity.
For general audio classification tasks (AS-20K, ESC-50), we observe that AudioMAE retains significantly more low-intensity tokens compared to AST. AudioMAE retains approximately 1.9× more tokens belonging to low-intensity clusters ($C_1$ and $C_2$) after pruning than AST in both datasets.
While both models and datasets retain empty (bottom-left most) patches after pruning, AudioMAE prunes all of those patches in the speaker identification task (VoxCeleb-1).

\begin{table}[ht]
  \caption{Results of discarding tokens clustered by intensity.}
  \label{tab:discard}
  \centering
  \setlength{\tabcolsep}{4pt} 
  \resizebox{0.95\linewidth}{!}{
   \begin{tabular}{c|cccc|ccc}
\toprule
\multirow{2}{*}{\textbf{block}} & \multicolumn{4}{c|}{\textbf{AudioMAE}} & \multicolumn{3}{c}{\textbf{AST}} \\
\cmidrule(lr){2-8}
 & \textbf{AS-20K} & \textbf{SPC-2} & \textbf{ESC-50} & \textbf{Vox-1} & \textbf{AS-20K} & \textbf{SPC-2} & \textbf{ESC-50} \\
\midrule
$\mathbf{L}/1$  & 35.9 & 97.43 & 91.54 & 94.39 & 35.5 & 95.36 & 88.56 \\
$\mathbf{L}/3$  & 35.8 & 97.62 & 91.24 & 94.62 & 36.6 & 96.56 & 92.49 \\
$\mathbf{L}/5$  & 36.0 & 97.75 & 91.86 & 94.68 & 37.9 & 97.22 & 94.69 \\
$\mathbf{L}/7$  & 36.4 & 97.91 & 92.26 & 94.91 & 38.4 & 97.31 & 94.83 \\
$\mathbf{L}/9$  & 36.5 & 98.10 & 92.75 & 95.21 & 38.6 & 97.35 & 94.93 \\
$\mathbf{L}/11$ & 37.2 & 98.25 & 93.05 & 95.62 & 38.7 & 97.32 & 94.98 \\
\midrule
$\mathbf{H}/1$  & 27.4 & 71.14 & 71.71 & 66.17 & 19.5 & 44.74 & 54.50 \\
$\mathbf{H}/3$  & 27.1 & 71.30 & 72.55 & 66.88 & 21.7 & 63.50 & 57.13 \\
$\mathbf{H}/5$  & 26.8 & 74.50 & 72.94 & 67.66 & 24.9 & 94.40 & 67.66 \\
$\mathbf{H}/7$  & 28.0 & 78.54 & 75.78 & 73.36 & 27.5 & 97.10 & 79.65 \\
$\mathbf{H}/9$  & 30.1 & 88.84 & 81.19 & 86.37 & 28.9 & 97.30 & 82.24 \\
$\mathbf{H}/11$ & 35.3 & 98.22 & 91.52 & 94.13 & 33.2 & 97.30 & 90.93 \\
\bottomrule
\end{tabular}
}
\end{table}

\section{Ablation Studies}

\subsection{Impact of Removing Tokens from Low or High Intensity Regions}
We assess the impact of discarding tokens belonging to low- (C1, C2: $\mathbf{L}$) or high-intensity (C4, C5: $\mathbf{H}$) groups on accuracy during inference. In this test, the tokens belonging to these groups are pruned after being processed by specific blocks. In Table~\ref{tab:discard}, $\mathbf{L}/i$ and $\mathbf{H}/i$ indicate the block index $i$ at which low- and high-intensity token groups are removed, respectively.
High-intensity tokens contribute significantly to model performance.
In addition, low-intensity tokens become more important for general audio classification tasks than for speech tasks.
By comparing the gaps in accuracy drops between low‑intensity and high‑intensity tokens, we see that high‑intensity tokens require deeper blocks to extract features for prediction. This also indicates that they contain richer information than low‑intensity tokens.
Overall, we could also observe that AudioMAE shows greater robustness to the loss of both token types across all datasets compared to AST.

\subsection{Using Different Feature Aggregation Methods}
\begin{figure}[ht]
  \centering
  \includegraphics[width=0.9\linewidth]{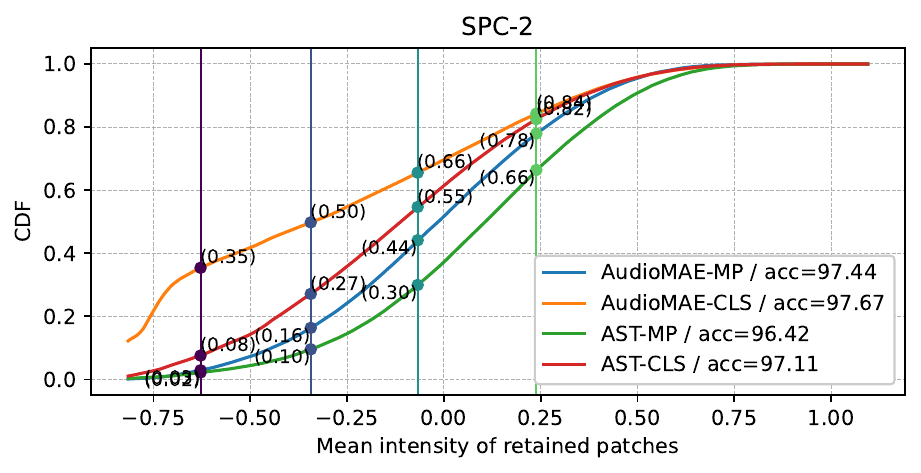}
  \includegraphics[width=0.9\linewidth]{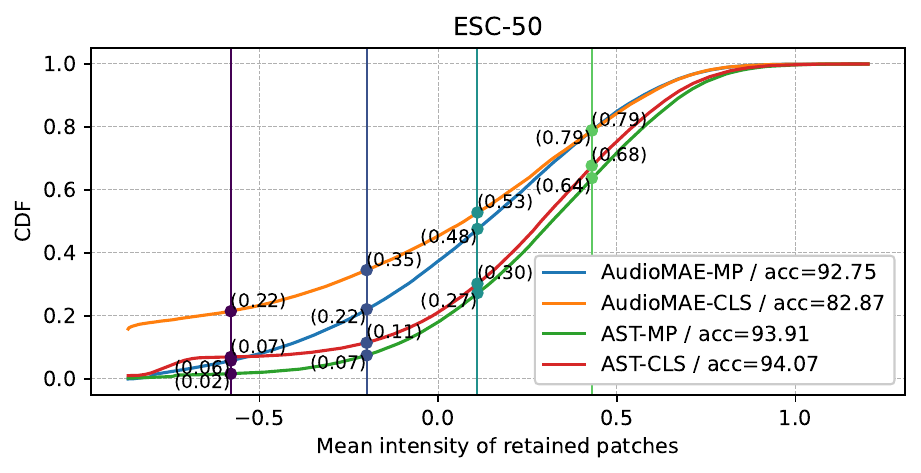}
  \caption{Cumulative distribution of retained tokens' mean-intensity of corresponding Mel-spectrogram patches. keep-rate is set to 0.5. Vertical color lines indicate the boundary of mean-intensity clusters.}
  
  \label{fig:cdf}
\end{figure}

We investigated whether different attention scoring and aggregation strategies—token-to-token attention with mean pooling versus [CLS] attention—affect the retention of low-intensity tokens, given that AudioMAE retains more such tokens than AST. We trained AudioMAE with a [CLS] token (AudioMAE-CLS) and AST with mean pooling (AST-MP) for classification. Fig.~\ref{fig:cdf} compares the cumulative distribution of retained tokens with respect to mean patch intensity. AudioMAE retained more low-intensity tokens than AST and achieved higher accuracy in SPC-2 task.
This difference can be explained by the models’ distinctive pretraining objectives: AudioMAE is trained in a self-supervised fashion to reconstruct every audio patch, regardless of its intensity. However, AST is trained in a supervised fashion, requiring the model to attend to tokens that are directly related to the label.
Although AudioMAE does not outperform AST on the ESC-50 task, it appears to learn better audio representations, as evidenced by its superior accuracy on VoxCeleb-1.

\section{Conclusion}
In this work, we show that TopK token pruning can be effectively applied to audio transformer models in classification tasks, achieving a competitive trade-off between accuracy and computational cost. We explore using intensity and variation of signals in patches as alternative token importance indicators, supported by positive Kendall correlations with attention scores. However, attention-based pruning consistently outperforms these statistics, suggesting that both low-intensity and low-variation patches are important. Visualization and ablation studies confirm that the model attends to such patches, especially in general audio classification tasks, highlighting their non-negligible contribution to higher classification accuracy.

\section{Discussions}
Other pruning methods such as DynamicViT, which uses MLP modules as token score predictor, may exhibit different pruning patterns. We believe that the effectiveness of token pruning varies across tasks due to two factors, which we leave for future investigation: (1) how uniquely an audio model produces token representations, and (2) the number of tokens a task requires to capture audio features. 


\bibliography{mybibfile}

\end{document}